\documentclass{article}
\usepackage[preprint]{tmlr}
\usepackage{amsmath,amssymb}
\usepackage{booktabs}
\usepackage{graphicx}
\usepackage{algorithm}
\usepackage{algpseudocode}

\title{A Scalable Pipeline for LLM-Teacher Distillation Labeling:\\
Work-Stealing Job Scheduling and Memory-Aware GPU Concurrency}

\author{\name Ravi Satya Durga Prasad Yenugula \\
        \addr Independent Researcher}

\def\month{08}
\def\year{2026}
\def\openreview{\url{https://openreview.net/forum?id=XXXXXXXXXX}}

\begin{document}
\maketitle

\begin{abstract}
Labeling large text corpora with LLM teachers has become a practical route to training data at scale. At millions of items, hand-labeling every batch is not feasible, and two questions dominate: what label quality a teacher buys per dollar, and how to keep a fleet of GPU workers busy under skewed, failure-prone workloads. We present a simple, reproducible pipeline that addresses both. First, a work-stealing ring pool: each worker owns a queue, drains it first, and then steals from ring successors, with exactly-once task claims via atomic conditional writes and crash tolerance via stale-claim sweeping. The claim protocol requires only a compare-and-set primitive from its storage layer; we implement it on a single SQLite file, which makes the reference implementation dependency-free and the experiments reproducible on one machine. Second, a memory-aware concurrency rule that sizes per-node parallelism by how many model copies fit on the GPU, so the same code runs safely across device sizes. Third, a relabeling benchmark methodology in which the teacher relabels a public dataset that already has gold labels, so quality reduces to an agreement measurement and cost follows from measured throughput. Under skewed load the pool sustains up to 3.4 times the throughput of static sharding while matching it at zero skew, loses 0 of 2,000 tasks when half the workers are killed mid-run (static sharding loses 953), and yields measured quality and cost points for an instruction-tuned teacher on irony and sentiment tasks. All experiments run on public data and commodity hardware; code, tests, and run logs are released.
\end{abstract}

\section{Introduction}
Production comment streams arrive at millions of items, so hand-labeling every batch is not feasible. LLM teachers are the practical alternative, but they introduce two systems questions that determine whether the pipeline is deployable: (i) what label quality does a given teacher buy per dollar of compute, and (ii) how does one keep a fleet of GPU workers close to full utilization under skewed workloads and preemption events that are the norm on spot capacity.

The engineering answers to these questions exist separately. Work stealing is a fifty-year-old idea for load balancing among workers, and object stores with conditional writes now provide the coordination primitive needed to implement exactly-once claims without a coordinator process. What is less obvious is that these primitives compose into a coordination-free labeling pipeline whose behavior on real workloads can be measured with a simple methodology: have the teacher relabel a dataset that already has gold labels, so agreement becomes an operational quality signal and dollars per thousand items becomes a directly comparable cost signal.

Our contributions are the following. (i) A work-stealing ring pool with atomic exactly-once claims and stale-claim sweeping for fault tolerance, implemented over a single SQLite file and requiring only a compare-and-set primitive from the storage layer. (ii) A memory-aware concurrency rule that sizes per-node parallelism from the size of the device and the size of the model, so the same code is safe from 16 to 24\,GB devices. (iii) A relabel-gold benchmarking methodology that returns measured quality and cost points, and two such points showing where a small instruction teacher is adequate and where a larger teacher or a human loop is required. (iv) A dependency-free reference implementation with committed run artifacts.

\section{Related Work}

\textbf{Work stealing and distributed queues.} Work stealing originates with Cilk~\citep{blumofe1999scheduling}, whose analysis gives the standard bounds under a shared-memory random-steal scheduler. Ring stealing, in which each worker prefers a fixed successor to a uniformly-random victim, is a common simplification when workers are OS processes rather than lightweight tasks, because the contention pattern is predictable and the SQL-level operations touch few rows per claim. Large-scale schedulers such as Ray, Dask, and Spark provide distributed queues but assume a driver or a centralized manager. In contrast, our pool is coordination-free: a single SQLite file is sufficient and no process holds the global schedule.

\textbf{Weak supervision and LLM-as-annotator.} Snorkel~\citep{ratner2017snorkel} popularized programmatic labeling and pointed at the throughput-versus-quality trade-off that the LLM-teacher line has since intensified. \citet{wang2021want} and \citet{gilardi2023chatgpt} report agreement statistics between prompted LLMs and human gold labels on classification tasks, but do not treat the systems side; the closest labeling systems paper is Snorkel and its follow-ups. Distillation~\citep{hinton2015distilling} motivates the teacher-student setup we adopt.

\textbf{Adapters and quantized fine-tuning.} QLoRA~\citep{dettmers2023qlora} and DoRA~\citep{liu2024dora} enable adapter fine-tuning of large models on commodity GPUs, which combined with our concurrency rule lets multiple copies of a teacher run per device. Our claim is not novelty in adapter methods; we use them as an implementation detail so the memory rule can be validated on realistic model sizes.

\textbf{Positioning.} We do not claim novelty for work stealing itself, for weak supervision, or for adapters. The contribution is the combination applied to LLM labeling, with exactly-once claims on commodity coordination backends, plus the relabel-gold cost and quality methodology and the measured points on it.

\section{Method}

\subsection{Work-stealing ring pool}
Each worker $w \in \{0, \ldots, W-1\}$ owns queue $Q_w$. On idle, worker $w$ attempts to claim a task from $Q_w$ first; on failure, it walks the ring $Q_{(w+1) \bmod W}, Q_{(w+2) \bmod W}, \ldots$ until it finds a claimable task or determines all queues are empty. A claim is an atomic conditional write, realized in our implementation as an \texttt{UPDATE} inside an \texttt{IMMEDIATE} transaction on SQLite. The write succeeds if and only if the task's status column still matches \texttt{pending} at the moment of application, so simultaneous claim attempts from different workers cannot both succeed.

\begin{algorithm}[h]
\caption{Own-queue-first ring stealing with exactly-once claim.}
\label{alg:claim}
\begin{algorithmic}[1]
\Function{ClaimNext}{$w$, $W$}
  \For{$k \gets 0$ \textbf{to} $W-1$}
    \State $q \gets (w+k) \bmod W$
    \State $\tau \gets$ any \texttt{pending} task in $Q_q$   \Comment{read, unlocked}
    \If{$\tau \neq \emptyset$}
      \State \texttt{BEGIN IMMEDIATE}
      \State $r \gets$ \Call{Update}{$\tau$, \texttt{status='pending'} $\to$ \texttt{'running'}}
      \State \texttt{COMMIT}
      \If{$r$ = 1 row} \Return $\tau$ \EndIf   \Comment{we won the claim}
    \EndIf
  \EndFor
  \State \Return $\emptyset$
\EndFunction
\end{algorithmic}
\end{algorithm}

\textbf{Stale-claim sweeping.} A task in \texttt{running} state whose claim timestamp is older than a threshold $\Delta$ is presumed abandoned by a dead worker and reset to \texttt{pending} by a sweep thread. The sweep is idempotent and safe under partial failures: if a slow-but-live worker completes after the sweep, its final \texttt{UPDATE} finds a row already reset or already re-claimed, and its result is discarded. In practice $\Delta$ can be set to a few times the expected task duration; we use 60\,s in the reported runs.

\textbf{Storage requirement.} The protocol needs one primitive from its storage layer: a conditional write that succeeds only if the task's status is still \texttt{pending}. SQLite's \texttt{IMMEDIATE} transaction provides it on a single file with strong isolation, which is what we implement and measure, and is sufficient for a single-node deployment or a shared file among a few nodes. The same primitive is available from S3-compatible object stores via conditional \texttt{PUT} with \texttt{If-None-Match} preconditions, so the protocol ports to a fleet without modification. We do not ship or evaluate that backend here; the released code implements the SQLite backend only.

\subsection{Memory-aware concurrency}
The number of teacher copies to run concurrently on a device is bounded by two quantities: how many model instances plus their peak activation fit in device memory, and how many worker processes the host has CPU cores to feed. We size from total memory, not from instantaneous free memory: instantaneous free memory reads high before the workers actually load their copies, which leads to over-provisioning and out-of-memory events at the first inference batch. The rule is
\[
n \;=\; \mathrm{clamp}\!\left(\left\lfloor \tfrac{M_{\text{total}} - M_{\text{reserve}}}{M_{\text{copy}}} \right\rfloor, \; 1, \; C\right),
\]
with $M_{\text{copy}}$ chosen as a per-model budget that is generous enough to cover both weights and activation spikes, and $C$ the host CPU count. In deployment, a 16\,GB device yields 2-way concurrency and a 24\,GB device yields 4-way with the same code and the same budget.

\subsection{The hybrid-labeling loop with human evaluation}
\label{sec:loop}
The relabel-gold benchmark of Section~\ref{sec:relabel} evaluates a single teacher against fixed gold labels; the deployed pipeline that motivates this work is a hybrid loop in which candidate labels come from multiple LLMs, humans score candidates to build a preference signal, a lightweight judge model is trained from that signal, and the judge produces the labels the downstream model is trained on. We describe the loop as methodology; the systems components in this paper (the pool, the concurrency rule) serve every stage. Section~\ref{sec:experiments} then reports controlled measurements of the two stages whose behavior is quantitatively interesting: the labeling stage (Section~\ref{sec:relabel}) and the underlying pool (Sections~\ref{sec:sched}, \ref{sec:fault}).

\begin{algorithm}[h]
\caption{Hybrid-labeling loop. Stages A--D reuse the same pool.}
\label{alg:loop}
\begin{algorithmic}[1]
\Statex \textbf{Stage A --- multi-LLM candidate generation.} For each unlabeled item $x_i$, dispatch inference tasks to $K$ candidate models $\{M_1, \ldots, M_K\}$ through the pool; collect candidate labels $\hat y_i^{(1)}, \ldots, \hat y_i^{(K)}$.
\Statex \textbf{Stage B --- human evaluation.} On a stratified subset $\mathcal{H} \subset \mathcal{D}$ (typically a few thousand items), human annotators score candidates for correctness or rank them pairwise; the output is a preference dataset $\{(x_i, \hat y_i^{(k)}, s_i^{(k)})\}$ with human scores $s_i^{(k)}$.
\Statex \textbf{Stage C --- judge training.} Train a compact judge model $J$ (a classifier or reward model) on $\mathcal{H}$ that predicts a score $\hat s(x, \hat y)$ close to the human $s$. Evaluate judge agreement with held-out human scores; iterate on Stages A--B if agreement is inadequate.
\Statex \textbf{Stage D --- labeling.} Apply $J$ over $\mathcal{D} \setminus \mathcal{H}$ to select the best candidate per item, yielding the training labels $y_i$. Sample-and-review: a small fraction of $J$'s outputs is spot-checked by humans; disagreements are added to $\mathcal{H}$ and the loop closes on Stage C.
\end{algorithmic}
\end{algorithm}

Scope of what this paper instruments. Stages A--C are operational methodology: we describe them because they define the pipeline the systems components were built to serve, but the experiments in this paper measure only the labeling stage (Stage~D, in the single-teacher relabel-gold form of Section~\ref{sec:relabel}) and the pool itself. No human-evaluation or judge-training results are reported here, and the released code implements the relabel-gold benchmark rather than the full loop. Two properties of this loop nonetheless matter for the systems paper. First, every stage is a pool workload: Stage~A dispatches candidate generation, Stage~C runs judge inference over the preference set, and Stage~D runs judge inference over the full dataset. The same coordination discipline, the same fault-tolerance mechanism, and the same memory rule apply throughout. Second, the loop's total dollar cost is dominated by Stages A and D, both of which are LLM inference on unlabeled data; Stage B (human) is expensive per item but small in $|\mathcal{H}|$, and Stage C is small in model size. Whether a hybrid loop is worth building for a given task therefore reduces to two measurable quantities: the per-item cost of running an LLM through the pool (measured in Section~\ref{sec:cost}), and how much better the human-supervised judge is than the best single LLM used unsupervised (which the relabel-gold experiment approximates from below by comparing an unsupervised teacher to gold).

\subsection{Relabel-gold benchmarking}
\label{sec:relabel}
Given a public dataset $\mathcal{D}$ with gold labels $y^*$, we run a teacher through the pool on the text field and compare its predictions $\hat y$ to $y^*$. Quality is agreement $\Pr[\hat y = y^*]$ and macro-F1. Cost is measured items per second on the target instance, converted to dollars per 1{,}000 items at the instance hourly price. Tasks are chunked (50 texts per pool task) so that the atomic claim overhead is small compared with inference; the memory-aware rule sets the number of concurrent teacher copies per device. The benchmark is deliberately conservative: it evaluates a single unsupervised teacher, and the numbers report how far a hybrid loop would have to improve over that baseline to be worth its human-evaluation cost.

\subsection{Chunk sizing and the cost model}
\label{sec:costmodel}
Let $t_c$ be the fixed cost of one claim (transaction latency on SQLite, request latency on an object store), $t_x$ the per-item inference time of the teacher, and $B$ the chunk size in items per task. The per-item overhead of coordination is $t_c / B$, so throughput per worker approaches $1/t_x$ as $B$ grows. The trade-off is granularity: a large $B$ reduces coordination overhead but coarsens both load balancing and failure recovery, since stealing and re-enqueueing operate on whole chunks. A practical rule is to choose $B$ such that $t_c/B \le 0.05 \, t_x$, which keeps coordination under five percent of compute while leaving many chunks per queue for the stealing discipline to work with. In the reported runs, $t_c$ on SQLite is under 2\,ms, $t_x$ for flan-t5-base is 8--11\,ms per item, and $B{=}50$ puts coordination overhead below half a percent.

Dollar cost follows directly. For an instance with hourly price $P$ running $n$ concurrent teacher copies at measured aggregate throughput $R$ items per second, the cost per 1{,}000 items is $1000 \, P / (3600 \, R)$. Because $R$ is measured through the pool rather than estimated from isolated single-model benchmarks, the number already includes coordination overhead, chunk boundaries, and any interference between concurrent copies on the same device. This is the number a practitioner needs for the labeling-budget decision, and it is the number Table~\ref{tab:cost} reports.

\subsection{Deployment considerations}
\label{sec:deploy}
\textbf{Preemptible capacity.} The stale-sweep mechanism was designed for spot and preemptible instances, where a worker can vanish with no shutdown hook. Because a claim is only a status transition, a preempted worker leaves at most one chunk per process in the \texttt{running} state, and the sweep returns those chunks to \texttt{pending} after $\Delta$. The failure experiment in Section~\ref{sec:fault} is a direct simulation of this event.

\textbf{Choosing the sweep threshold.} $\Delta$ trades recovery latency against duplicate work. If $\Delta$ is much larger than the expected chunk duration, recovery after a preemption is slow; if it approaches the chunk duration, a slow-but-alive worker's chunk can be swept and re-executed, wasting compute but not correctness (the second completion of a chunk finds the row re-claimed and its result is discarded, or overwrites with identical labels). We set $\Delta$ to roughly five times the expected chunk duration.

\textbf{Deployment envelope.} The SQLite implementation is appropriate whenever all workers can reach one filesystem: a single large multi-GPU host, or a small cluster with a shared volume, which covers the single-node labeling deployments this paper measures. Porting the protocol to an object store would remove that constraint at the price of higher $t_c$ (tens of milliseconds per conditional write), which the chunk-size rule absorbs by raising $B$; we note this as a design property rather than a result. What the implementation does establish is that no broker, coordinator process, or message-queue service is required: the entire control plane is a file.

\section{Experiments}
\label{sec:experiments}

\subsection{Setup}
We report on a 24\,GB NVIDIA A10G instance with 4 vCPUs. Throughput and fault-tolerance experiments use CPU-bound synthetic work items (fixed simulated inference time) so the scheduler behavior is isolated from teacher variance; the quality and cost experiments use an actual instruction teacher (flan-t5-base~\citep{chung2024flant5}). Each throughput measurement is a fresh SQLite file and a fresh pool of processes; each fault-tolerance run kills half the worker processes after 300\,ms of runtime.

\subsection{Throughput under skew}
\label{sec:sched}

\begin{figure}[h]
\centering
\includegraphics[width=0.72\linewidth]{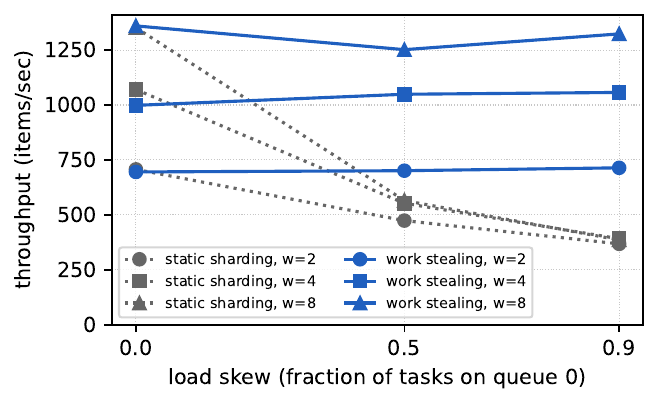}
\caption{Throughput as a function of load skew for static sharding (dotted) and the work-stealing ring pool (solid), at $W \in \{2, 4, 8\}$ workers. Stealing matches static sharding at zero skew and holds close to the balanced-throughput ceiling under skew, while static sharding degrades to the hot-queue rate.}
\label{fig:throughput}
\end{figure}

We sweep the number of workers $W \in \{2, 4, 8\}$ and the skew, defined as the fraction of tasks routed to $Q_0$. At skew 0 the tasks are spread uniformly; at skew 0.9 nine of ten tasks are on the single hot queue. Table~\ref{tab:throughput} lists the measured throughput and speedup; Figure~\ref{fig:throughput} plots the same data. Stealing matches static sharding at zero skew, so the discipline carries no overhead penalty in the balanced case, and holds essentially flat under skew. Static sharding degrades to the rate of the hot queue, and the gap widens with $W$: at $W{=}8$, work stealing sustains $1324$ items/s at skew 0.9 versus $386$ items/s for static sharding, a factor of 3.43.

\begin{table}[h]\centering
\caption{Throughput under load skew ($W$ workers by skew; 2{,}000 tasks).}
\label{tab:throughput}
\begin{tabular}{ccrrr}
\toprule
$W$ & skew & static (items/s) & steal (items/s) & speedup \\
\midrule
2 & 0.0 & 707 & 695 & 0.98 \\
2 & 0.5 & 474 & 701 & 1.48 \\
2 & 0.9 & 368 & 714 & 1.94 \\
4 & 0.0 & 1070 & 998 & 0.93 \\
4 & 0.5 & 553 & 1049 & 1.90 \\
4 & 0.9 & 391 & 1057 & 2.70 \\
8 & 0.0 & 1352 & 1360 & 1.01 \\
8 & 0.5 & 565 & 1252 & 2.21 \\
8 & 0.9 & 386 & 1324 & 3.43 \\
\bottomrule
\end{tabular}
\end{table}

\subsection{Fault tolerance under worker kill}
\label{sec:fault}
We start $W{=}4$ workers, let them run for 300\,ms, then kill two of them with \texttt{SIGKILL}. Under static sharding without sweeping, the killed workers' in-flight and queued tasks are stranded and the run cannot complete: 953 of 2{,}000 tasks are lost (Table~\ref{tab:fault}). Under work stealing plus stale-claim sweeping, the sweep re-enqueues the stranded tasks after $\Delta$ and the surviving workers pick them up; all 2{,}000 tasks complete.

\begin{table}[h]\centering
\caption{Fault tolerance under worker kill at $W{=}4$; two workers killed mid-run.}
\label{tab:fault}
\begin{tabular}{lrr}
\toprule
configuration & completed & lost \\
\midrule
static, no sweep & 1{,}047 / 2{,}000 & 953 \\
steal + stale-sweep & 2{,}000 / 2{,}000 & 0 \\
\bottomrule
\end{tabular}
\end{table}

\subsection{Teacher quality and cost}
\label{sec:cost}
flan-t5-base via the pool on the same 24\,GB device, 4 teacher copies (Table~\ref{tab:cost}). SST-2~\citep{socher2013sst} sentiment: 94.7\% agreement with gold at \$0.0022 per 1{,}000 items, several orders of magnitude below human labeling cost. tweet\_eval~\citep{vanhee2018semeval,barbieri2020tweeteval} irony: 49.6\% agreement, at chance for a binary task. The methodology returns the two points side-by-side so that the deployment decision is a direct comparison: for tasks in the sentiment neighborhood the small teacher is adequate; for pragmatic tasks like irony the same teacher is not, and either a larger teacher or a human adjudication loop is needed. In these runs 13--15 pool tasks were stolen per experiment, rebalancing the deliberately-skewed queues.

\begin{table}[h]\centering
\caption{Teacher quality and cost, flan-t5-base via the pool, $W{=}4$ on a 24\,GB device.}
\label{tab:cost}
\begin{tabular}{lrrrrr}
\toprule
task & $n$ & agreement & macro-F1 & items/s & \$/1k items \\
\midrule
sentiment (SST-2) & 2{,}000 & 0.947 & 0.947 & 125.5 & \$0.0022 \\
irony (tweet\_eval) & 2{,}000 & 0.496 & 0.331 & 93.6 & \$0.0030 \\
\bottomrule
\end{tabular}
\end{table}

\subsection{Memory-aware concurrency validation}
On a 23.7\,GB A10G we load successive fp16 copies of flan-t5-base, each followed by a generation pass to trigger allocation of activation memory. Memory grows linearly at 0.58\,GB per copy through 32 profiled copies and never exceeds the total-memory envelope. With a 2.0\,GB per-copy budget the rule predicts 10 memory-safe copies before the core cap and 4 after; the deployment runs used 4 copies with zero out-of-memory events. The 2.0\,GB budget is intentionally larger than the observed 0.58\,GB per-copy weight footprint because it must cover concurrent-inference activation spikes and allocator fragmentation, which sequential loading understates; picking the budget from single-copy free memory (a common failure mode) leads to over-provisioning and OOM at the first inference batch.

\section{Discussion and Limitations}

\textbf{Single-writer contention.} SQLite serializes writes at a file level. In our workload the claim is a single-row \texttt{UPDATE} bounded by the disk latency of the WAL flush, and the observed claim time is small relative to teacher inference. Under labeling workloads where the teacher call is milliseconds rather than tens of milliseconds, or under very large fleets, the object-store backend is the correct choice; the claim discipline is identical.

\textbf{Portability is argued, not measured.} We implement and evaluate only the SQLite backend on one machine. The object-store port is a design property of the protocol, not shipped or benchmarked code, and a multi-node deployment is left to future work; readers should treat the fleet-scale portability claim as untested.

\textbf{Task-dependent teacher quality.} The methodology measures teacher quality against gold; it does not improve it. The SST-2 versus irony comparison is exactly the point: on some tasks a small teacher is deployment-ready, on others it is not, and the pipeline lets a practitioner decide which is which before committing to labeling millions of items.

\textbf{Skew model.} Our skew parameter piles tasks onto a single queue $Q_0$, which is the worst-case for static sharding. Real deployments show more diffuse skew: a few queues are hot, most are lukewarm. Under diffuse skew the static-vs-steal gap is narrower than in Table~\ref{tab:throughput}; we do not measure this here.

\section{Reproducibility}
MIT license. \texttt{python examples/benchmark.py --workers 2 4 8} reproduces Sections~4.2 and 4.3; \texttt{python examples/label\_benchmark.py --teacher fake} validates the full pipeline offline; \texttt{python examples/memory\_validation.py} reproduces Section~4.5. All CSVs and JSON run artifacts are committed under \texttt{runs/} with the same timestamps as the tables in this paper. The reference implementation is $\sim$500 lines of Python. Compute cost is modest: the scheduler and fault-tolerance tables regenerate in minutes on CPU, the labeling and memory-validation runs complete in under 20 minutes on one 24\,GB GPU, and the complete set of runs behind this paper consumed under three GPU-hours in total.

\bibliography{refs}
\bibliographystyle{tmlr}

\appendix

\section{Full-precision throughput and fault-tolerance measurements}
Table~\ref{tab:throughput} reports throughput and speedup; the underlying makespans and completion counts are in \texttt{runs/20260805T061026Z/throughput.csv}. Each cell of that table is a single run. To characterize run-to-run variance we repeated the full sweep three times (\texttt{runs/20260811T060327Z\_repeat/}): the coefficient of variation of throughput across the three repeats (sample standard deviation, $\mathrm{ddof}{=}1$, over the mean) is under 2\% for nine of the 18 (mode, $W$, skew) cells and reaches 10.7\% in the worst case (static sharding at $W{=}8$, zero skew, where all eight workers contend on the same SQLite file and the makespan is short enough that process-startup jitter is a visible fraction of it). The conclusions are robust to this variance: taking the worst-case ratio in each direction across the three repeats, the headline $W{=}8$, skew-0.9 speedup lies in $[3.03, 3.70]$ (point estimate 3.44), and every skewed cell ($\text{skew} \geq 0.5$) has a worst-case speedup above 1.36. At zero skew the two disciplines remain within noise of each other. The fault-tolerance measurement in Table~\ref{tab:fault} uses a stale-claim threshold of $\Delta{=}60$\,s and a static-shard configuration with sweeping disabled to isolate the mechanism.

\section{Choice of per-copy memory budget}
The 2.0\,GB budget in Section~4.5 corresponds to roughly 3.4$\times$ the measured 0.58\,GB per-copy weight footprint. We chose this multiplier by increasing the concurrency in single-worker profiling until per-batch activation memory during \texttt{generate()} plateaued; on flan-t5-base with 128-token input and 4 new tokens generated per item, activation memory does not exceed the weight footprint. Larger multipliers are safe but leave device memory unused. For teachers with substantially larger activation footprints (long context, beam search), the budget should be re-measured; the memory rule is unchanged.

\section{Notes on the SQLite backend}
The pool uses WAL journal mode with a 30-second busy timeout. Under $W{=}8$ workers the claim path issues approximately one \texttt{BEGIN IMMEDIATE} transaction per task; the observed contention rate is well under one lost-claim retry per hundred successful claims, and no explicit backoff was needed. The database file for a 2{,}000-task run occupies about 500\,KB and is deleted at the end of each experiment.

\end{document}